\documentclass[aps,prl,twocolumn,showpacs,groupedaddress]{revtex4}  
\usepackage{graphicx}  
\usepackage{dcolumn}   
\usepackage{bm}        
\usepackage{epsfig}
\usepackage{color}
\usepackage{amssymb}
\usepackage{amsmath}
\usepackage{wasysym}
\usepackage{pifont}

\definecolor{bleu}{rgb}{0,0,1}
\definecolor{vert}{rgb}{0,0.5,0}
\definecolor{rouge}{rgb}{1,0,0}
\definecolor{rose}{rgb}{0.9,0.3,0.7}
\definecolor{azur}{rgb}{0,0.5,0.5}
\definecolor{orange}{rgb}{1,0.5,0.2}

\newcommand{\be}{\begin{equation}}
\newcommand{\ee}{\end{equation}}
\newcommand{\ba}{\begin{eqnarray}}
\newcommand{\ea}{\end{eqnarray}}

\newcommand{\ie}{\textit{i.e.}~}

\newcommand{\eg}{\textit{e.g.}~}
\newcommand{\leg}[1]{{\it #1}}

\newcommand{\mathnotation}[2]{\newcommand{#1}{\ensuremath{#2}}}
\mathnotation{\ldef}{\mathrel{\raisebox{.069ex}{:}\!\!=}}
\mathnotation{\rdef}{\mathrel{=\!\!\raisebox{.069ex}{:}}}

\mathnotation{\pd}{\partial}           
\mathnotation{\dint}{\,{\mathrm{d}}}   

\mathnotation{\br}{\textbf{r}}
\mathnotation{\ri}{\textbf{r}_i}	
\mathnotation{\x}{x_i}
\mathnotation{\y}{y_i}
\mathnotation{\Dr}{\Delta \textbf{r}_i^t\left(\tau\right)}
\mathnotation{\Dx}{u_i^t\left(\tau\right)}
\mathnotation{\Dy}{v_i^t\left(\tau\right)}
\mathnotation{\deltaDx}{\delta u_i^t\left(\tau\right)}
\mathnotation{\deltaDy}{\delta v_i^t\left(\tau\right)}
\mathnotation{\sigx}{\sigma_x\left(\tau\right)}
\mathnotation{\sigy}{\sigma_y\left(\tau\right)}
\mathnotation{\sig}{\sigma_{\phi}\left(\tau\right)}
\mathnotation{\nuu}{\frac{d\log\sigma}{d\log\tau}}	
\mathnotation{\Kux}{\kappa_x\left(\tau\right)}
\mathnotation{\Kuy}{\kappa_y\left(\tau\right)}
\mathnotation{\Gene}{\mathcal{F}\left(\lambda,\tau\right)}
\mathnotation{\Genx}{\mathcal{F}_{\tau}\!\left(x\right)}

\mathnotation{\q}{q_i^t\left(a,\tau\right)}
\mathnotation{\Qt}{Q^t\left(a,\tau\right)}
\mathnotation{\Qbar}{\bar{Q}\left(a,\tau\right)}
\mathnotation{\Xifour}{\chi_4\left(a,\tau\right)}
\mathnotation{\Ssix}{\mathcal{S}_6\left(a,\tau\right)}
\mathnotation{\Gfour}{G_4\left(r;a,\tau\right)}	
\mathnotation{\Cfour}{C_4\left(T;a,\tau\right)}

\begin{document}


\title{Super-diffusion around the rigidity transition: \\
L\'evy and the Lilliputians}

\author{F.~Lechenault}
\affiliation{CEA Saclay/SPCSI, Grp. Complex Systems \& Fracture, F-91191 Gif-sur-Yvette, France}
\author{R.~Candelier}
\affiliation{LPS, Ecole Normale Sup\'erieure, URA D 1306, 24 rue Lhomond 75005 Paris, France.}
\author{O.~Dauchot}
\affiliation{CEA Saclay/SPEC, URA2464, L'Orme des Merisiers, 91 191 Gif-sur-Yvette, France}
\author{J.-P.~Bouchaud}
\affiliation{Capital Fund Management, 6-8 Bd Haussmann, 75009 Paris, France}
\author{G.~Biroli}
\affiliation{CEA Saclay/IPhT, UMR2306, L'Orme des Merisiers, 91 191 Gif-sur-Yvette, France }

\date{\today}

\begin{abstract}
By analyzing the displacement statistics of an assembly of horizontally vibrated bidisperse frictional grains in the vicinity of the jamming transition experimentally studied before~\cite{Lech1}, we establish that their superdiffusive motion is a genuine L\'evy flight, but with `jump' size very small compared to the diameter of the grains. The vibration induces a broad distribution of jumps that are random in time, but correlated in space, and that can be interpreted as micro-crack events at all scales. As the volume fraction departs from the critical jamming density, this distribution is truncated at a smaller and smaller jump size, inducing a crossover towards standard diffusive motion at long times. This interpretation contrasts with the idea of temporally persistent, spatially correlated currents and raises new issues regarding the analysis of the dynamics in terms of vibrational modes.
\end{abstract}

\maketitle 


\section{Introduction}

As the volume fraction of hard grains is increased beyond a certain point, the system jams and is able to sustain mechanical stresses. This rigidity/jamming transition has recently been experimentally investigated~\cite{Lech1,Lech2} in an assembly of horizontally vibrated bi-disperse hard disks, using a quench protocol that produces very dense configurations, with packing fractions beyond the glass density $\phi_g$, such that the {\it structural} relaxation time $\tau_\alpha$ is much larger than the experimental time scale. There is a density range $\phi_g < \phi < \phi_a$ where the strong vibration can still induce micro-rearrangements through collective contact slips that lead to partial exploration of the portion of phase space restricted to a particular frozen structure. For $\phi_g < \phi < \phi_J \approx 0.842$, the system is frozen but not rigid; the system can only sustain an external stress for  $\phi$ larger than $\phi_J$, which appears as a genuine critical point where a dynamical correlation length and a correlation time simultaneously diverge, showing that the dynamics occurs by involving progressively more collective rearrangements. 

One of the most surprising result of~\cite{Lech1} was the discovery of a superdiffusive regime in the vicinity of $\phi_J$. In a time range that diverges when $\phi \to \phi_J$, the typical displacement of the grains grows as $\tau^\nu$, with $\nu > 1/2$, \ie faster than the familiar diffusive $\sqrt{\tau}$ law, while always remaining small compared to the diameter of the grains! It was also found that dynamical heterogeneities reach a maximum when $\tau=\tau^*(\phi)$, precisely when the superdiffusive character of the motion is strongest~\cite{Lech2}, in agreement with the general bound on the dynamical susceptibility established in~\cite{science,JCP1}. This superdiffusion was interpreted as the existence of large scale convective currents, which were tentatively associated to the extended {\it soft modes} that appear when the system loses or acquires rigidity at $\phi_J$~\cite{Wyart}, and along which the system should fail. A similar interpretation of the dynamics of particulate systems close to the glass or jamming transitions was promoted in~\cite{Harrowell}. The intuitive idea is that energy barriers that the system has to cross to equilibrate are essentially in the directions (in phase space) defined by the soft modes. Within this picture, the {\it harmonic} motion of the particles around a metastable configurations can be used to guess the structure of the {\it anharmonic} barrier crossing events. This is certainly reasonable when the barriers are small, so that the top of the barrier is still in a quasi-harmonic regime. 

\begin{figure}[!t]
\centering
\includegraphics[width=0.60\columnwidth]{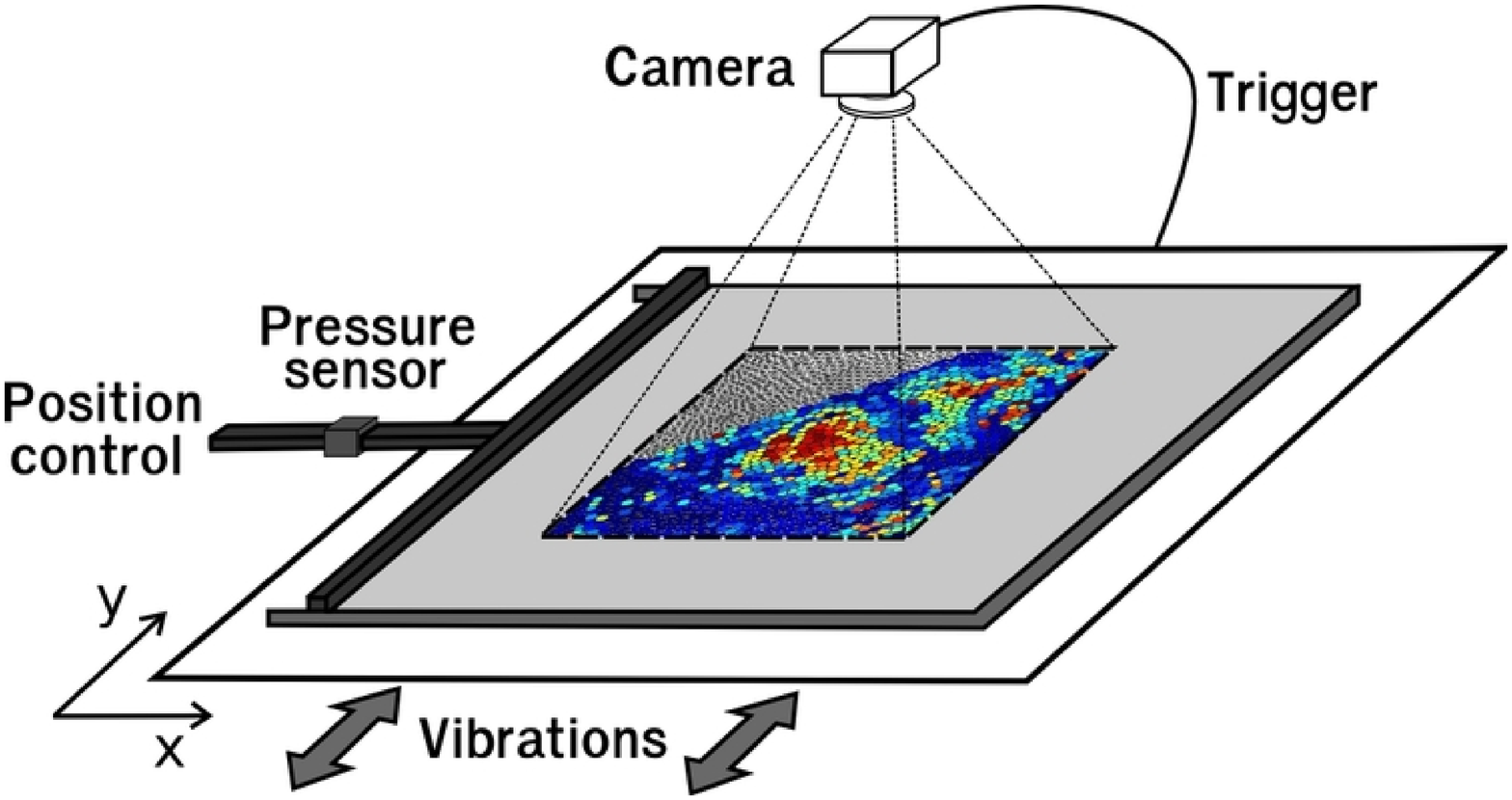}
\includegraphics[width=0.35\columnwidth]{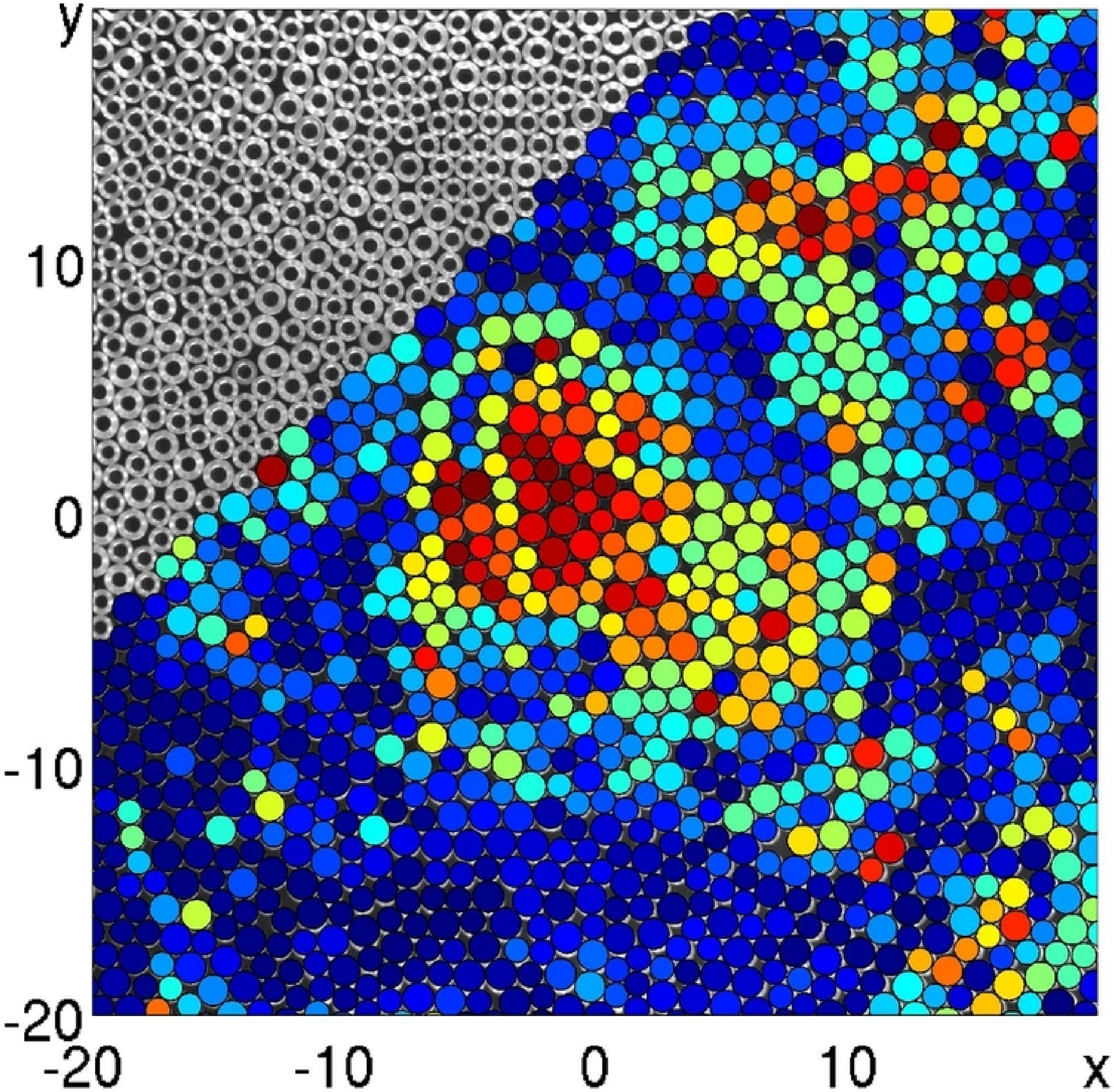}
\caption{
\leg{Left:} Sketch of the experimental setup. 
\leg{Right:} Picture of the grains together with an illustration of the relaxation pattern of the field $q_i$ at $\tau^*$ (see text below for details).}
\label{fig:setup}
\end{figure}

The primary aim of the present paper is to revisit the above interpretation for our system of hard-disks with friction, in the light of a deeper analysis of the statistics of displacements. To our surprise, we find that the distribution of rescaled displacements is, in the superdiffusive regime, accurately given by an isotropic L\'evy stable distribution $L_\mu$. Furthermore, the exponent $\mu$ characterizing this L\'evy stable distribution is equal to $1/\nu$, where $\nu$ is the superdiffusion exponent, precisely what one would expect if the motion of the grains was a {\it L\'evy flight}, \ie a sum of uncorrelated individual displacements with a power-law tail distribution of sizes such that the variance of the distribution diverges \cite{Levy,BG}. This divergence only occurs at $\phi=\phi_J$, but is truncated away from the critical point, which explains why the motion reverts to normal diffusion at very large times. This finding shows that the rearrangements corresponding to the maximum of dynamical correlations cannot be thought of as large scale currents that remain coherent over a long time scale $\tau^*$. Superdiffusion is {\it not} induced by long-range temporal correlations of the velocity field, as was surmised in~\cite{Lech1}. Quite on the contrary, the total displacement on scale $\tau^*$ is made of a large number of {\it temporally incoherent} jumps with a broad distribution of jump sizes. As is well known, a L\'evy flight is, over any time $\tau$, dominated by a handful of particularly large events. Hence, the only predictability of the total displacement between $t=0$ and $t=\tau^*$ on the basis of the motion in any early period, say between $t = 0 \to \tau^*/10$, comes from the presence of one of these large jumps in $[0,\tau^*/10]$. The situation for frictional grains might therefore be quite different from what happens in supercooled liquids or frictionless hard spheres, when soft modes should indeed play an important role.

On the other hand, the very fact that individual jumps are broadly distributed in such a dense system where grains can hardly move means that these large jumps are necessarily collective, with a direct relation between the size of the jump and the number of particles involved that we discuss below. The physical connection between superdiffusion and dynamical heterogeneities in this system becomes quite transparent. Note however that although we speak about ``broad distributions'' and ``large jumps'', one should bear in mind that all of this takes place on minuscule displacement scales, a few $10^{-2}$ of the grains diameter! Our L\'evy flights are therefore Lilliputian walks
with a diverging second moment...

The experimental set-up and the quench protocols are described in detail in~\cite{Lech1}, and summarized in Fig.~\ref{fig:setup}. The stroboscopic motion of a set of 8500 brass cylinders (``grains'') is recorded by a digital video camera. The cylinders have diameters $d_{1} = 4\pm0.01 mm$ and $d_{2} = 5\pm0.01 mm$ and are laid out on a horizontal glass plate that harmonically oscillates in one direction at a frequency of 10$~Hz$ and with a peak-to-peak amplitude of 10$~mm$. The cell has width $L \approx 100$ $d_{small}$, and its length can be adjusted by a lateral mobile wall controlled by a $\mu m$ accuracy translation stage, which allows us to vary the packing fraction of the grains assembly by tiny amounts ($\delta \phi/\phi \sim 5 \, 10^{-4})$. The position of the grains is tracked within an accuracy of $2.\, 10^{-3} d_{1}$. In the following, lengths are measured in $d_{1}$ units and time in cycle units.

\section{Statistical characterization of the motion of grains}

\begin{figure}[t!]
\centering
\begin{minipage}[t]{.99\linewidth}
\includegraphics[width=0.55\columnwidth]{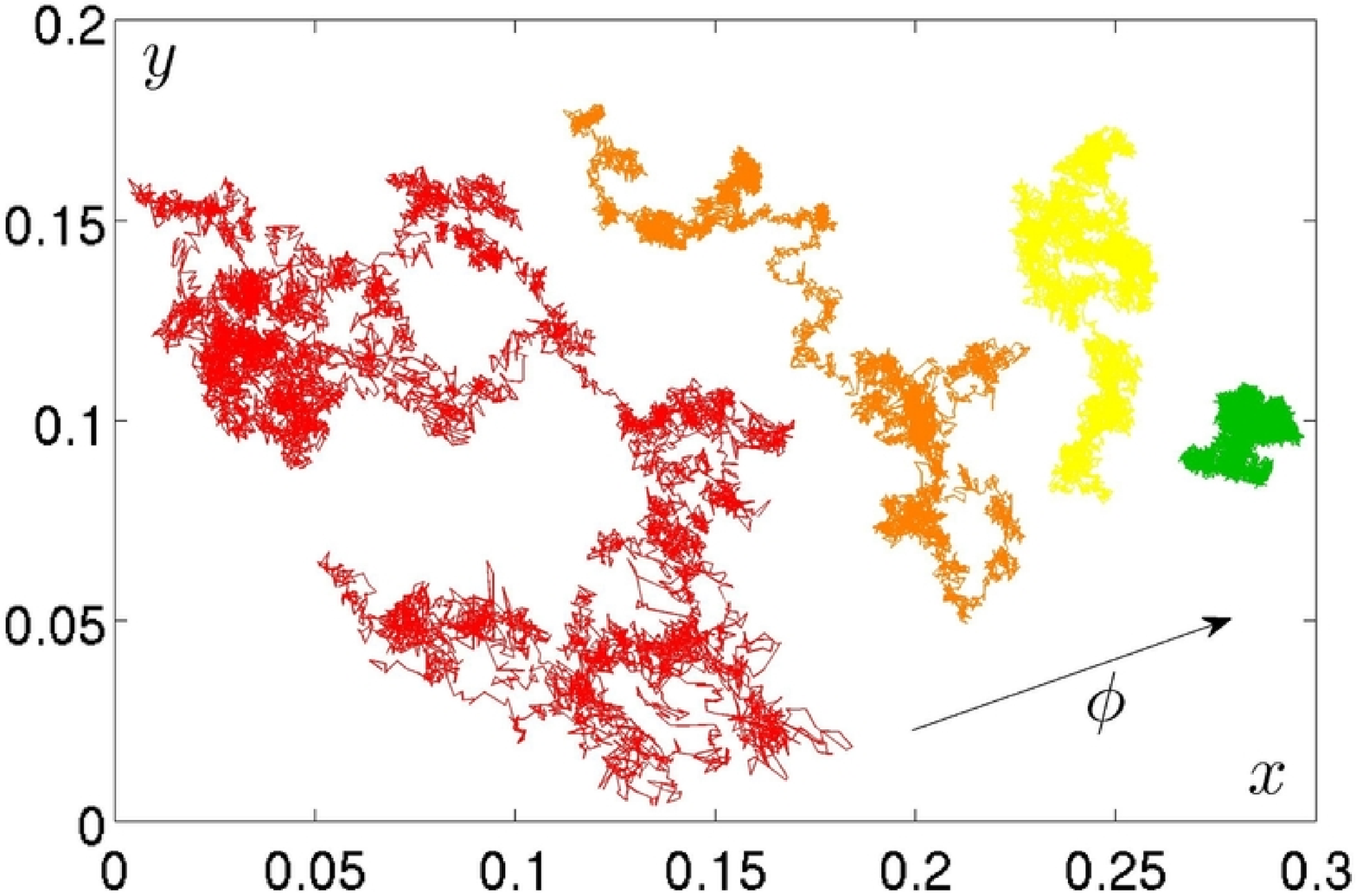}
\includegraphics[width=0.40\columnwidth]{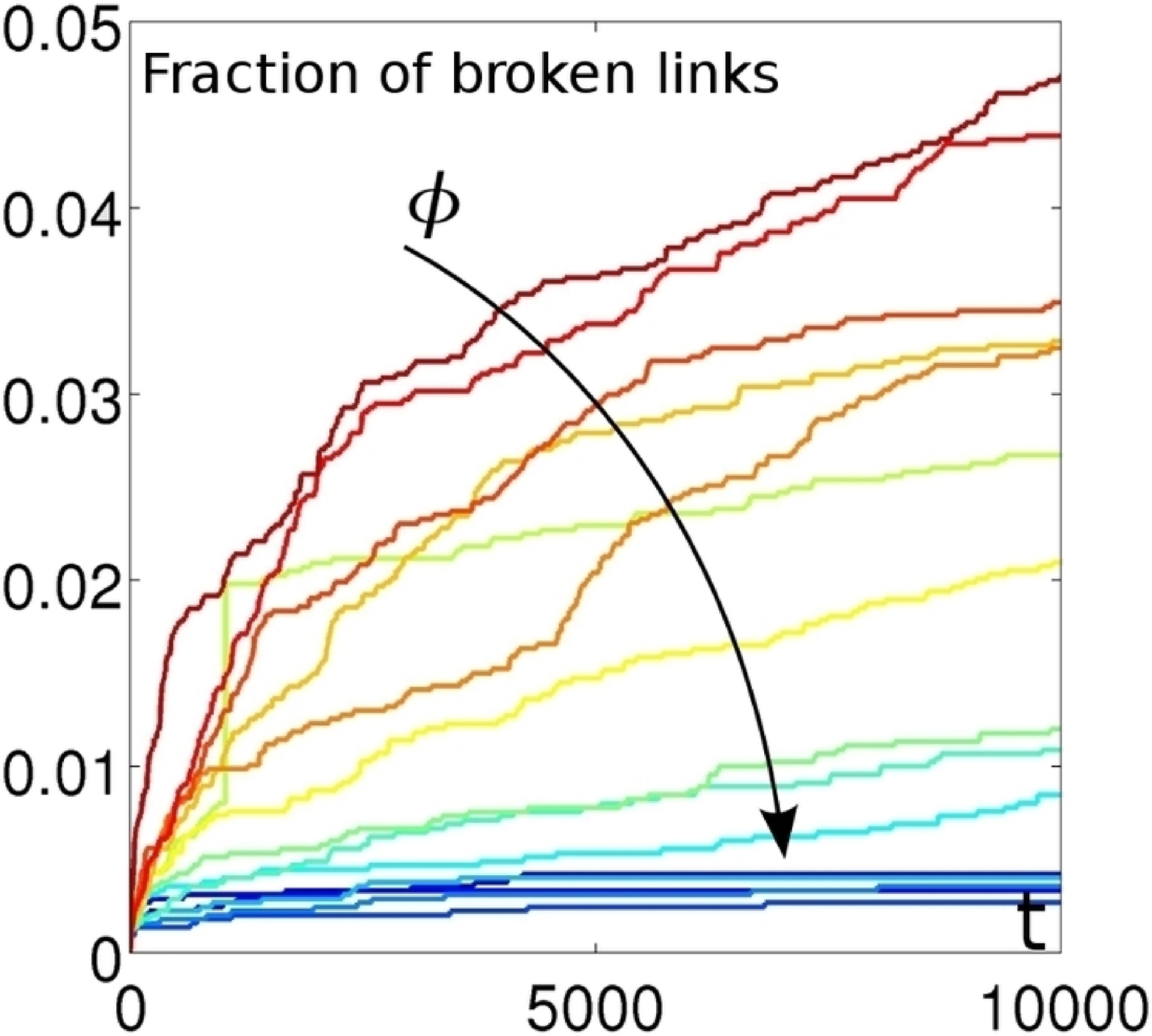}
\end{minipage}
\caption{
\leg{Left}: Typical trajectories of a grain during a 10000 cycle acquisition. The packing fractions are $\phi= 0.8402$ (red), $0.8417$ (orange), $0.8422$ (yellow), $0.8430$ (green). $\phi_J$ would stand between the orange and yellow trajectories.
\leg{Right}: Fraction of broken neighborhood relations (links in the Vorono\"i tessellation sense), as a function of time for all packing fractions.
}
\label{fig:traj}
\end{figure}

Fig.~\ref{fig:traj}-left shows the trajectories of a single grain during the whole experimental window for packing fractions sitting on both sides of the transition $\phi_J$. As expected, the typical displacements are much larger at lower packing fractions. The so-called caging dynamics appears clearly on the trajectories obtained at larger packing fractions: the grains seem to stay confined around fixed positions for a long time and then hop around from one cage to another over longer time scales. The striking feature of these curves is the remarkably small amplitude of the motion that we alluded to above. Even at the lowest packing fractions, the grains do not move much further than a fraction of a diameter over the total course of one experimental run.  Another evidence of the absence of structural relaxation is given by the low percentage of neighboring links (in the Vorono\"i sense) that are broken during the $10^4$ cycles of an experimental run (see Fig.~\ref{fig:traj}-right). This fraction goes from around $5\%$ for the loosest packing fraction to around $0.2\%$ for the densest. This confirms that for densities around $\phi_J$, the system is deep in the glass phase where structural relaxation is absent, \ie in a very different regime from the one studied in \cite{Marty,Abate,Candelier}.

We will denote by $\textbf{R}_i^t=\left(x_i^t,y_i^t\right)$ the vector position of grain $i$ at time $t$ in the center-of-mass frame. Lagged 
displacements are defined as $\textbf{r}_i^t(\tau) = \textbf{R}_i^{t+\tau} - \textbf{R}_i^t$ for grain $i$ between time $t$ and $t+\tau$. We have checked in details that the statistics of 
$\textbf{r}_i^t$ along the $x$ and the $y$ axes are identical, or more precisely that the motion is isotropic, in spite of the strongly anisotropic nature of the external drive. This in itself is a non trivial observation, that shows that the random structure of the packing is enough to convert a directional large scale forcing into an isotropic noise on small scales. From now on, we will thus focus on the total displacement $r_i^t = |\textbf{r}_i^t|$. We measure typical displacements for a density $\phi$ and lag $\tau$ as the mean absolute displacement, defined as:
\be
\sigma_{\phi}\left(\tau\right) \equiv \left\langle |\textbf{r}_i^t(\tau)| \right\rangle_{i,t}
\ee
where the time average $\langle \cdot \rangle_t$ is performed over $10,000$ cycles. We have chosen to estimate $\sigma_{\phi}(\tau)$ by the mean absolute value deviation instead of the root mean square displacement because -- as we shall see -- the later diverges when approaching the transition.

\begin{figure}[!t]
\centering
\begin{minipage}[t]{.99\linewidth}
\includegraphics[width=0.49\columnwidth]{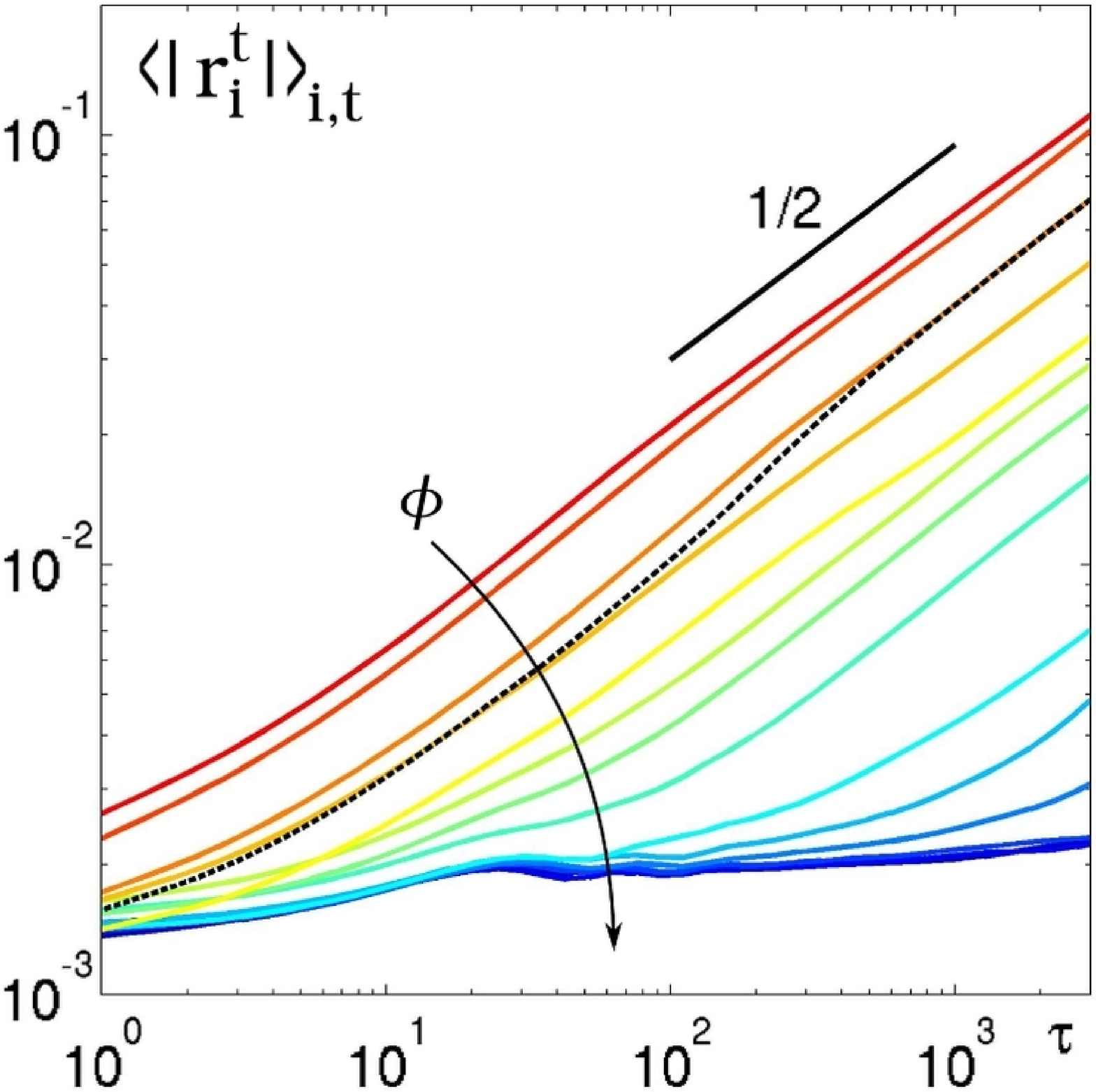}
\includegraphics[width=0.49\columnwidth]{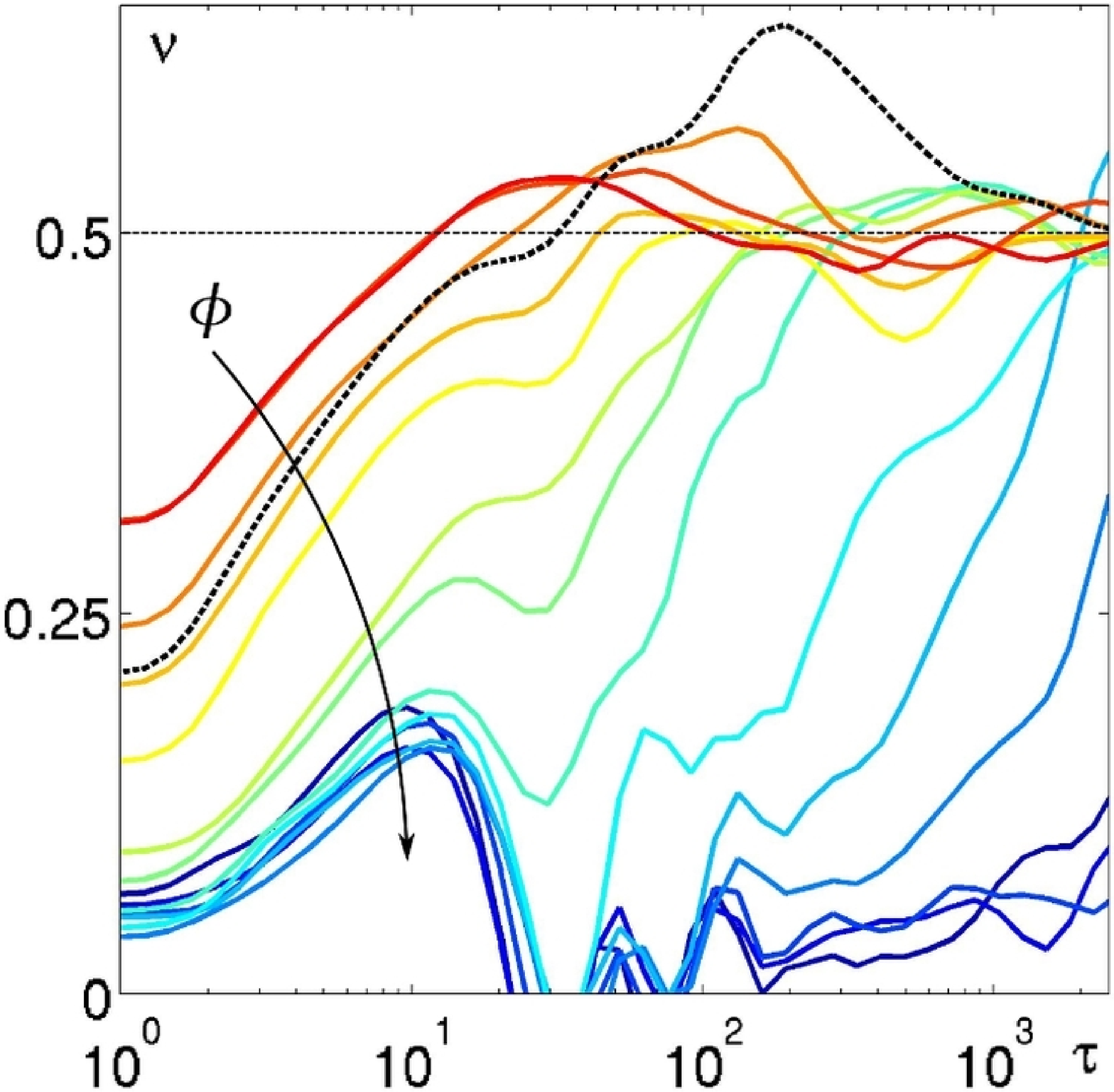}
\includegraphics[width=0.49\columnwidth]{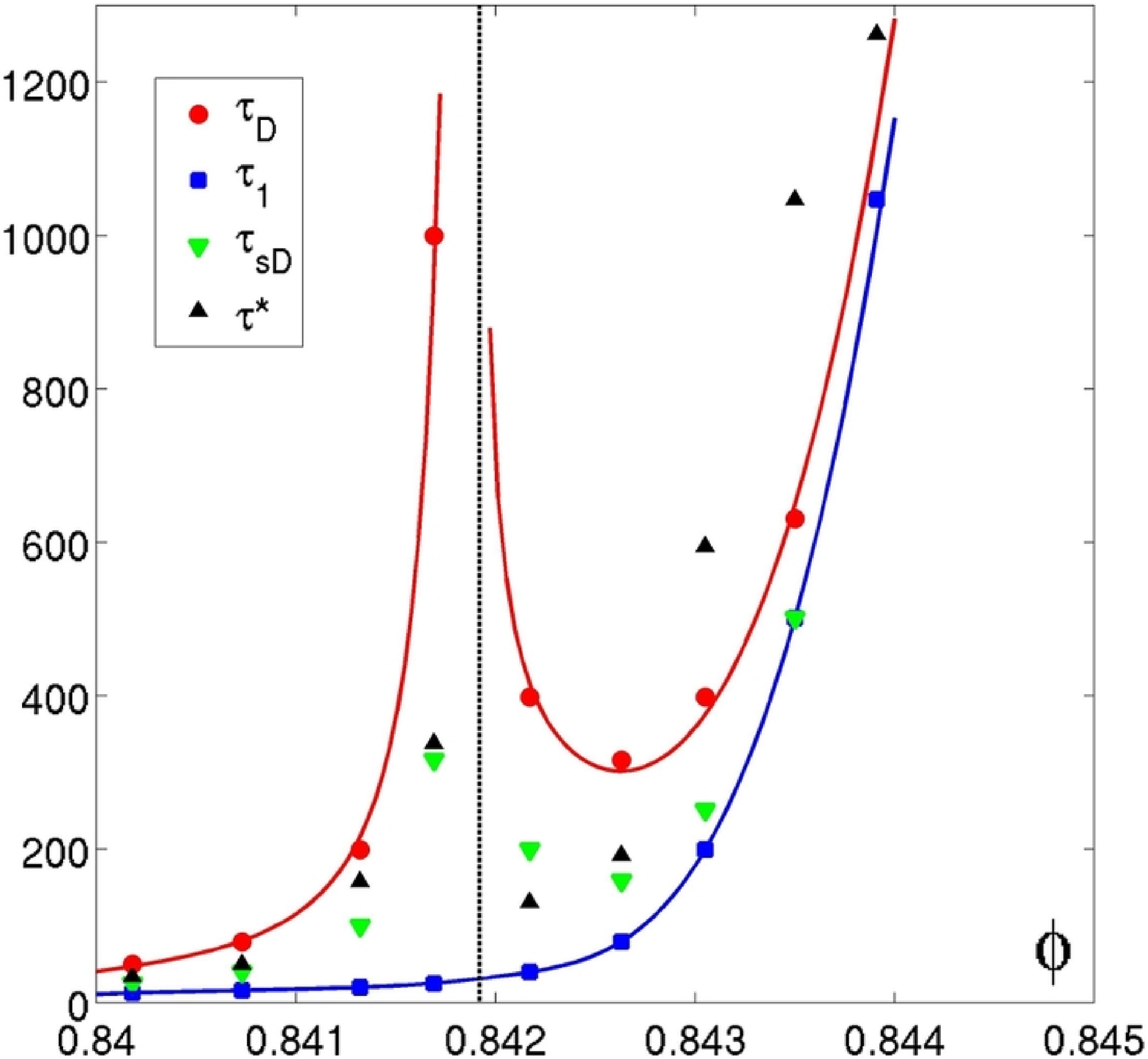}
\includegraphics[width=0.49\columnwidth]{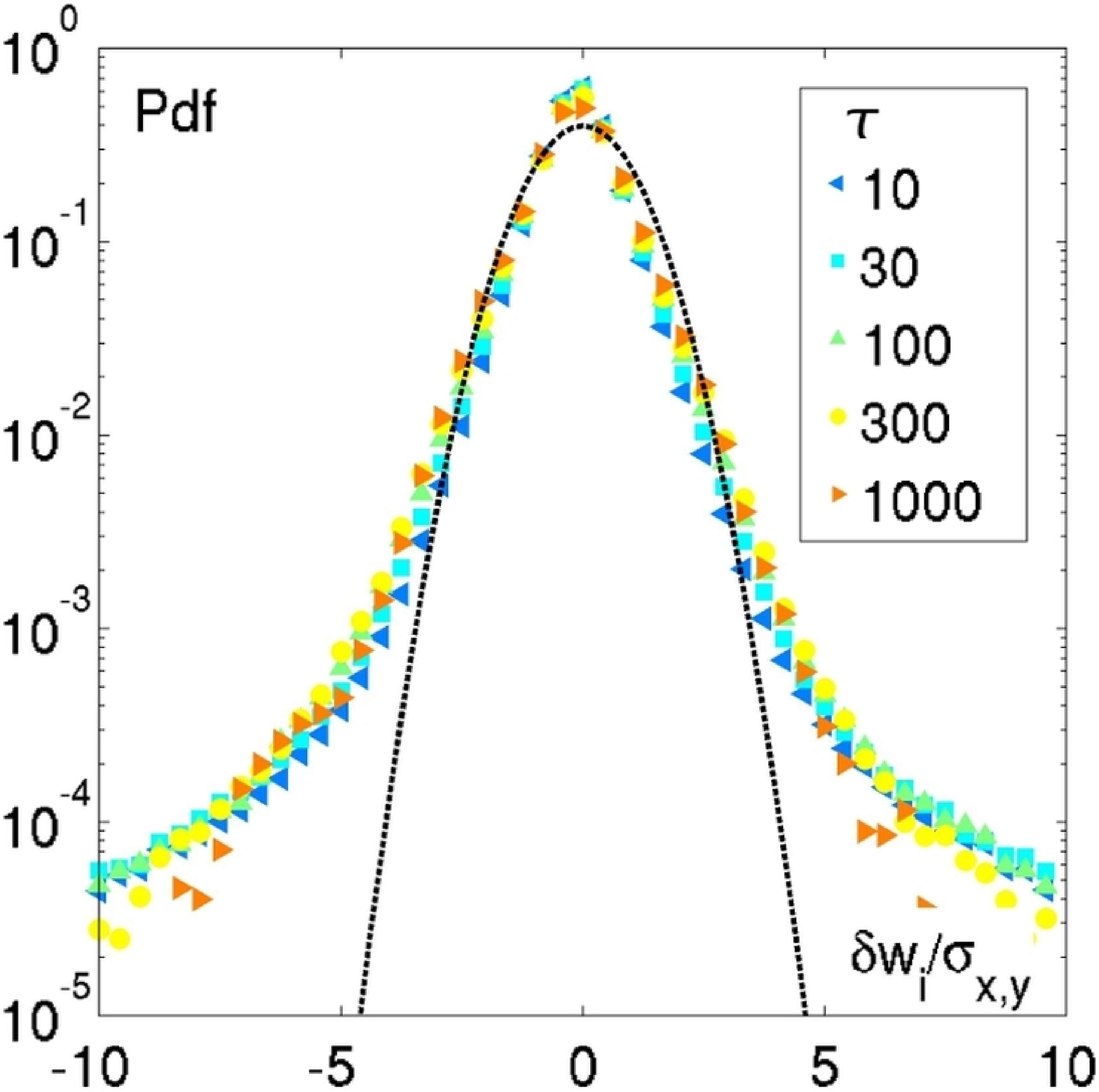}
\end{minipage}
\caption{
\leg{Top:} Average displacement $\sigma_{\phi}\left(\tau\right) \equiv \langle |\textbf{r}_i^t| \rangle_{i,t}$ (left); and local slope $\nu = {\partial \ln \sigma_{\phi}\left(\tau\right)}/{\partial \ln \tau}$ (right) as functions of the lag $\tau$ for all studied values of $\phi$. The dotted dark line corresponds to $\phi_J$.
\leg{Bottom left:} Characteristic time scales extracted from the analysis of $\nu(\tau)$ including $\tau^*$, which will be introduced 
and discussed in the section devoted to dynamical heterogeneities.
\leg{Bottom left:} Pdf of the displacements $r_i$, $P_\tau(r)$, normalized by $\sigma_{\phi}(\tau)$, for different values of the lag time 
$\tau$ at $\phi=0.8446$ (right). A Gaussian distribution (black dotted curve) is shown for comparison.}
\label{fig:rms-dv}
\end{figure}

The evolution of $\sigma_{\phi}(\tau)$ with the packing fraction is presented in Fig.~\ref{fig:rms-dv}, see~\cite{Lech1}. Note the very small values of $\sigma_{\phi}(\tau) \ll d_1$ at all timescales, which is a further indication that the packing indeed remains in a given structural arrangement. At low packing fractions $\phi < \phi_J$ and at small $\tau$, $\sigma_{\phi}(\tau)$ displays a sub-diffusive behavior. At longer time, diffusive motion is recovered. As the packing fraction is increased, the typical lag at which this cross-over occurs becomes larger and, at first sight, $\sigma_{\phi}(\tau)$ does not seem to exhibit any special feature for $\phi=\phi_J$ (corresponding to the bold line in Fig.\ref{fig:rms-dv}). Above $\phi_J$, an intermediate plateau appears before diffusion resumes. However, a closer inspection of $\sigma_{\phi}(\tau)$ reveals an intriguing behavior, that appears more clearly on the local logarithmic slope $\nu={\partial \ln \sigma_{\phi}(\tau)}/{\partial \ln(\tau)}$ shown in Fig.~\ref{fig:rms-dv}. When $\nu=\frac{1}{2}$, the motion is diffusive, whereas at small times, $\nu < \frac{1}{2}$ indicates sub-diffusive behavior. At intermediate packing fractions, instead of reaching $\frac{1}{2}$ from below, $\nu$ overshoots and reaches values $\approx 0.65$ before reverting to $\frac{1}{2}$ from above at long times. Physically, this means that after the sub-diffusive regime commonly observed in glassy systems, the particles become {\it super-diffusive} at intermediate times before eventually entering the long time diffusive regime. At higher packing fractions, this unusual intermediate superdiffusion disappears: one only observes the standard crossover between a plateau regime at early times and diffusion at long times. In order to characterize these different regimes, we define three characteristic times: $\tau_1(\phi)$ as the lag at which $\nu(\tau)$ first reaches $1/2$, corresponding to the beginning of the super-diffusive regime, $\tau_{sD}(\phi)$ when $\nu(\tau)$ reaches a maximum $\nu^{*}(\phi)$ (peak of super-diffusive regime), and $\tau_{D}(\phi)$ where $\nu(\tau)$ has an inflection point, beyond which the system approaches the diffusive regime. These characteristic time scales are plotted as a function of the packing fraction in the bottom-left panel of Fig.~\ref{fig:rms-dv}. Whereas $\tau_1$ does not exhibit any special feature across $\phi_J$, both $\tau_{sD}$ and $\tau_{D}$ are strongly peaked at $\phi_J$. We have also shown on the same graph the time scale $\tau^*$ where dynamical heterogeneities are
strongest (see~\cite{Lech1} and below). As noted in the introduction, $\tau^*$ and $\tau_{sD}$ are very close to one another and we will identify these two time scales in the following. 

We now turn to the probability distribution of the displacements $r_i^t\left(\tau\right)$. We have represented on the bottom-right panel of Fig.~\ref{fig:rms-dv} the distributions accumulated for all grains and instants for the packing fractions closest to $\phi_J$ and several values of lag-time $\tau$. The horizontal axis has been normalized by the root mean square displacements, and a unit Gaussian is also plotted for comparison. We find that the tails of the distributions are much fatter than the Gaussian, unveiling the existence of extremely large displacements (compared to typical values) in the region of time lags corresponding to superdiffusion. As we will discuss below, these fat tails tend to disappear when one leaves the superdiffusion regime, \ie  when $|\phi - \phi_c|$ and/or $|\ln \tau/\tau^*|$ become large. 

Before characterizing more precisely these probability distributions, we would like to mention that we have removed ``rattling events'' from our analysis, \ie events where particles make an anomalously large back and forth motion, of amplitude larger than (say) $0.1 d_1$. Some of these events are due to the same grain rattling during a large fraction of the experimental run, while other events are localized in time for a given grain, and could actually well be part of the extreme tail of the distribution seen in Fig.~\ref{fig:rms-dv}. Our point here is that the observed fat-tails are not an artifact due to a few ``loose'' grains -- each and everyone of the grains seem to contribute at one point or another in its history to these fat tails.

A neat way to characterize the probability distribution of the displacements is to study the generating function of the squared displacements:
\be
\Gene\equiv\langle e^{-\lambda {r}_i^{t\,2}(\tau) }\rangle_{i,t} = \sum_{k=0}^\infty \frac{\left(-\lambda\right)^k}{k!} \langle {r}_i^{t\,2k}(\tau) \rangle.
\ee
As a benchmark, one can easily compute $\Gene$ in the case of isotropic Gaussian diffusion. One obtains: 
\be
\Gene_{G}=\tilde{\mathcal{F}}_{G}(x)=\frac{1}{1+x},
\ee
where $x=2 \sigma_{\phi}(\tau)^2 \lambda$ is the scaled variable. When computing $\Gene$ for the empirical data, one finds, as expected, systematic discrepancies with the $\Gene_{G}$. The rescaling of the different curves when plotted as a function of $x$, on the other hand, works very well, as can be inferred from the scaling of the probability distributions themselves. 

\begin{figure}[!t]
\centering
\begin{minipage}[t]{.99\linewidth}
\includegraphics[width=0.49\columnwidth]{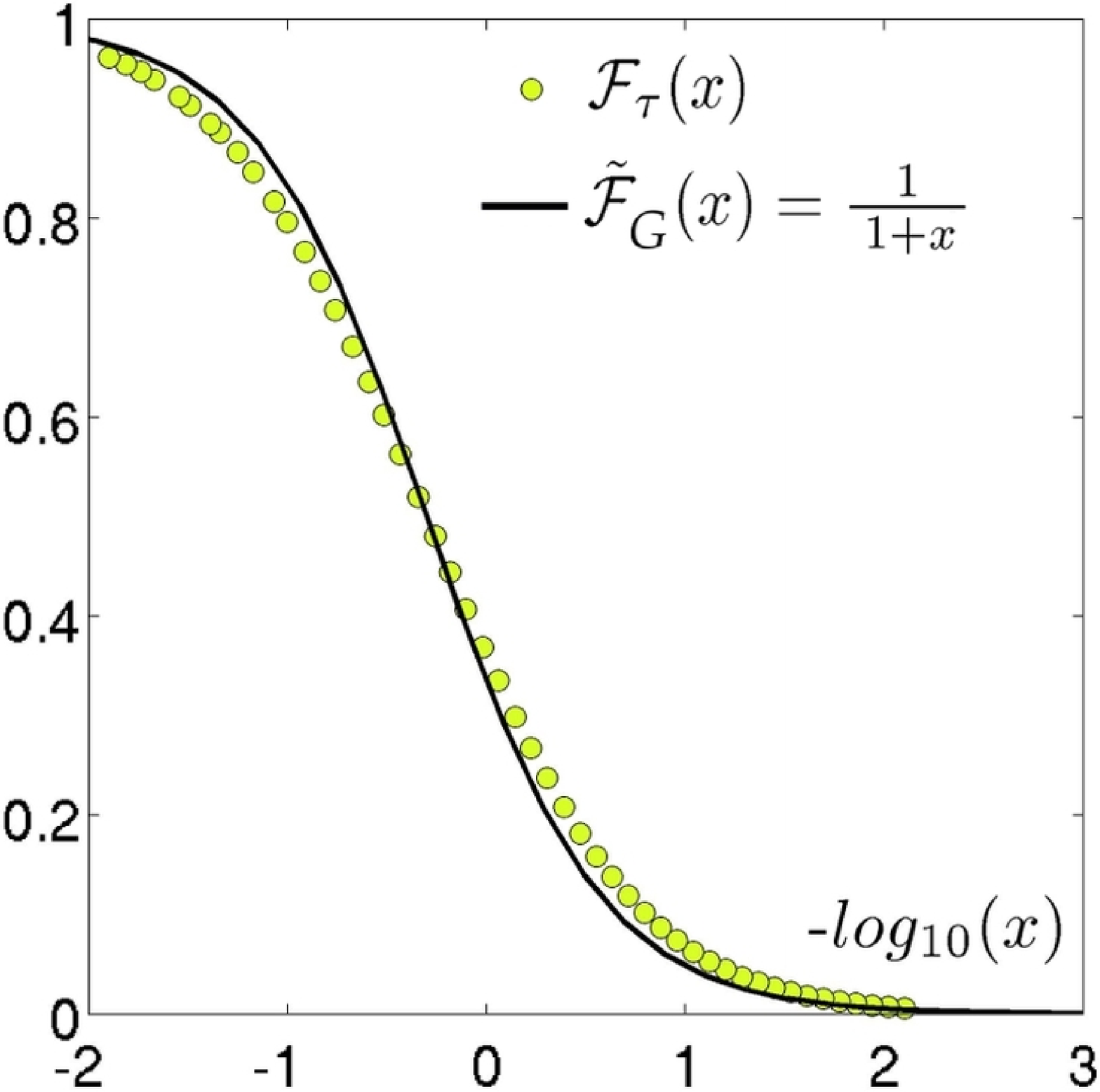}
\includegraphics[width=0.49\columnwidth]{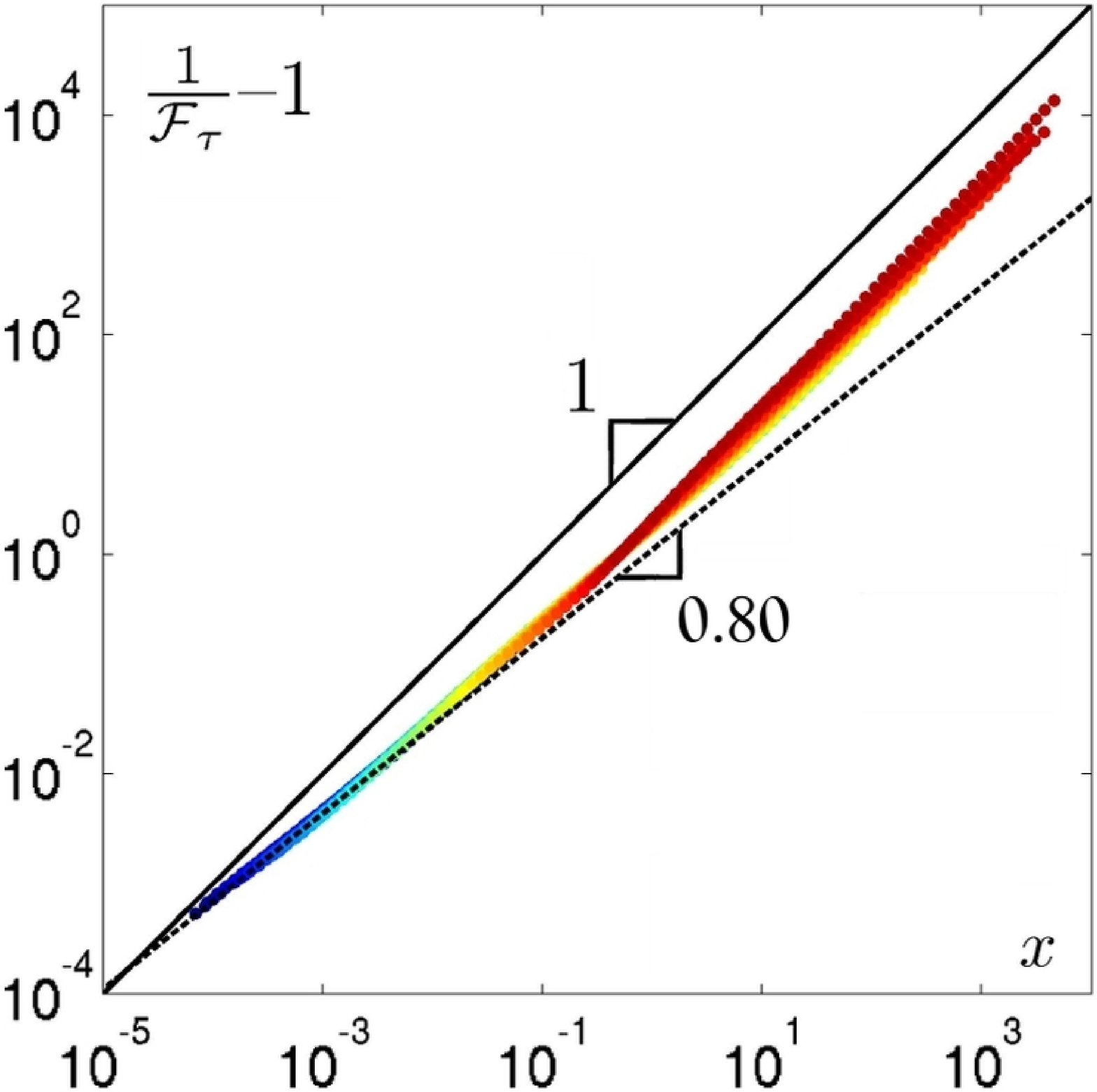}
\includegraphics[width=0.49\columnwidth]{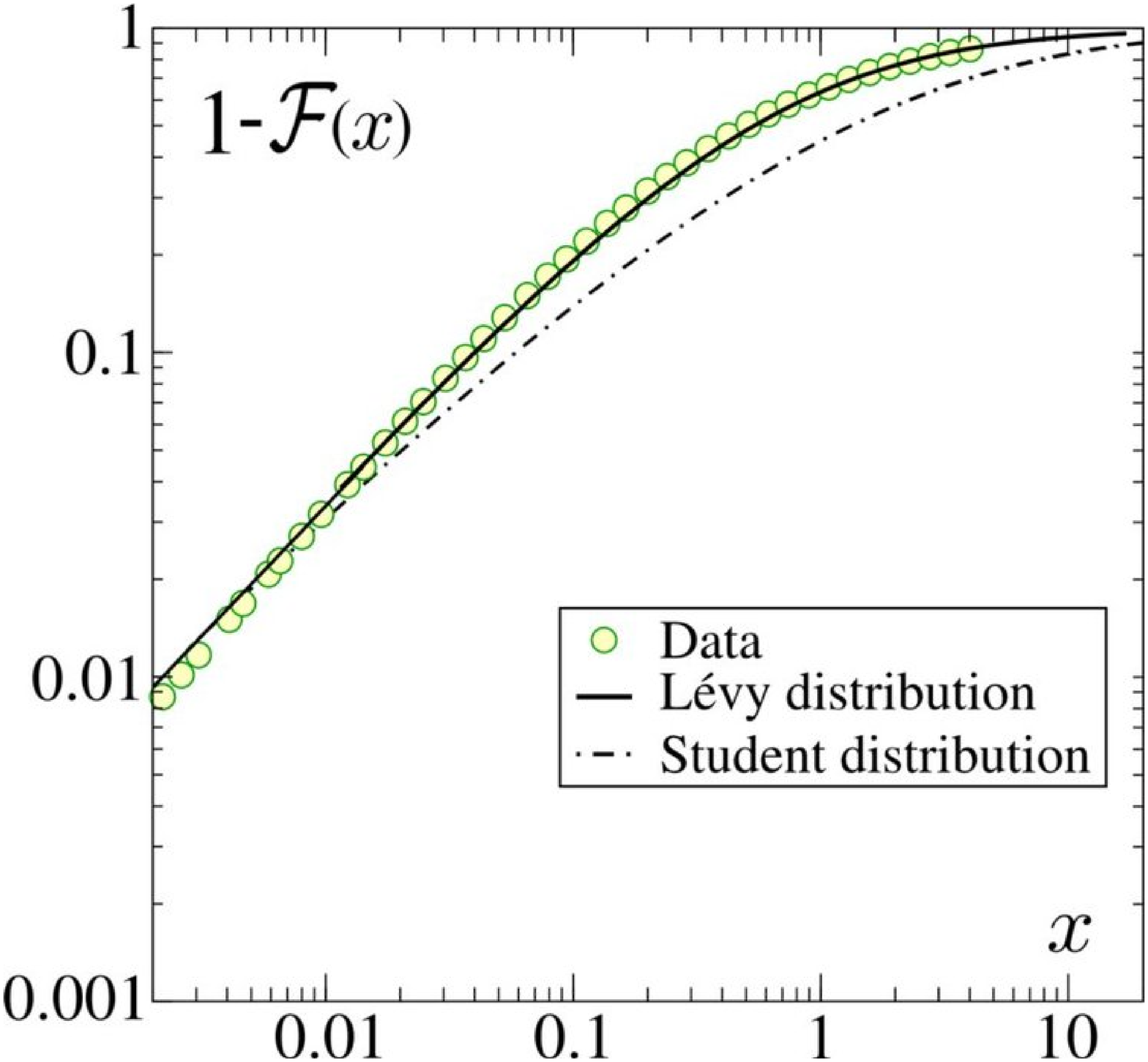}
\includegraphics[width=0.49\columnwidth]{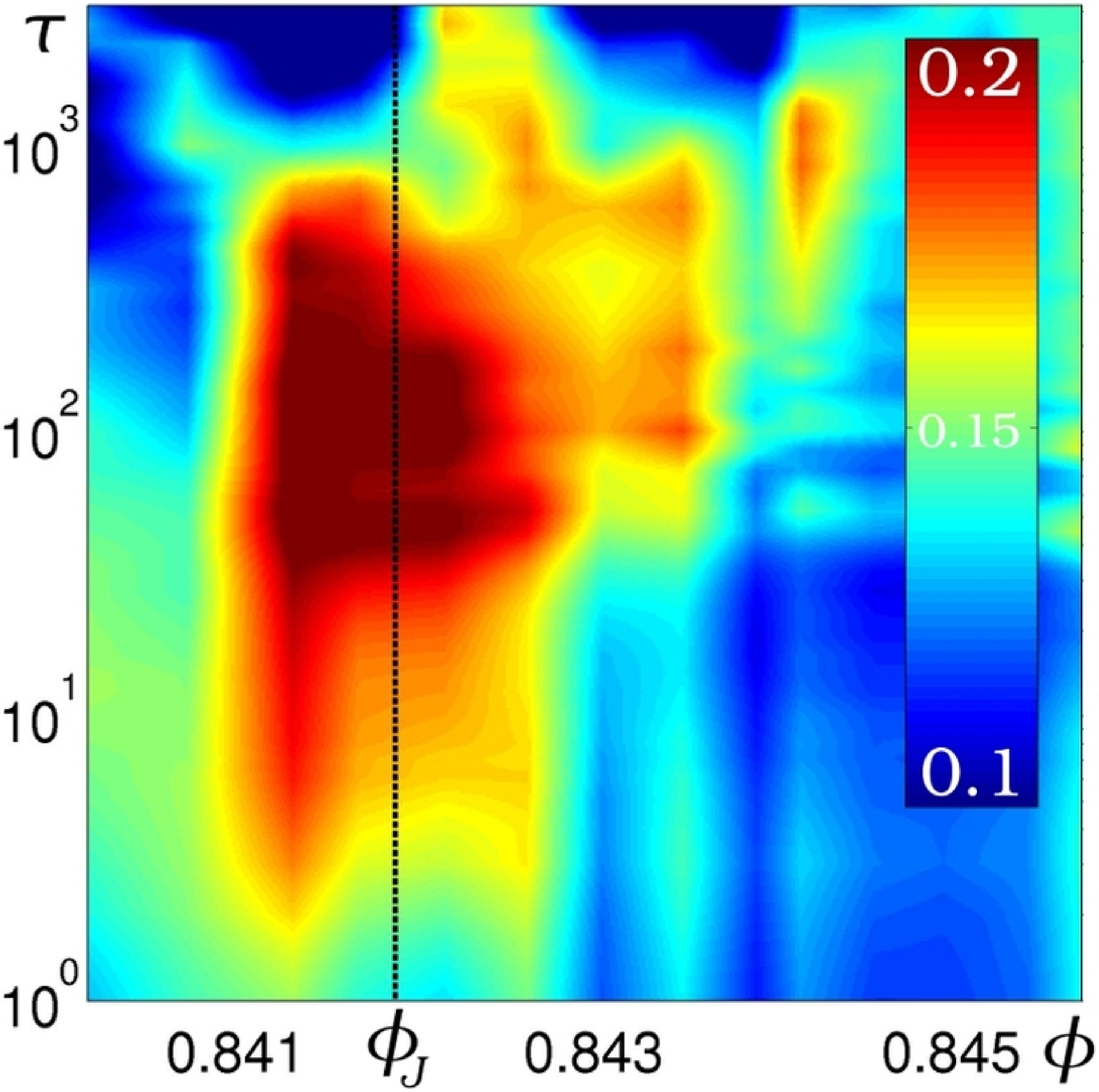}
\end{minipage}
\caption{
\leg{Top:} $\Genx$ as a function of $-\log_{10}(x)$ for $\tau=\tau^*=373$ (dots, left panel) and $1/\Genx-1$ as a function of $x$ for different values of $\tau$ (dots, right panel), from $1$ (blue) to $10^4$ (red). For both plots: $\phi=\phi_J$ and the solid line corresponds to the Gaussian fit $1/(1+x)$.
\leg{Bottom-left:} Fit of $\Genx$ for $\phi \approx \phi_J$ and $\tau \approx \tau^*$ using a L\'evy distribution model $\Gene_{L}$ with index $\mu=1.6$, and comparison with a Student distribution 
model $\Gene_{L}$ with the same small $x$ behavior.
\leg{Bottom-right:} Fitting parameter $\varepsilon=1-\mu/2$ as a function of $\phi$ and $\tau$.
}
\label{fig:gene}
\end{figure}

The most important finding is that the small $x$ behavior of the empirical $\Gene$ is singular: as shown in the top-right panel of Fig.~\ref{fig:gene}, $1/\Gene - 1$ behaves as $x^\alpha$ when $x \ll 1$, with $\alpha \approx 0.80$. It is easy to show that such a singular behavior is tantamount to the existence of a power-law tail in the distribution of $r(\tau)$, $P_\tau(r) \sim_{r \to \infty} r^{-1-\mu}$ with $\alpha=\min(1,\mu/2)$. The value of $\alpha$ therefore corresponds to $\mu \approx 1.6 < 2$, which means that the variance of the distribution of $r$ is formally divergent, or at least dominated by a large physical cut-off $r_{\max}$. This suggests to fit $\Gene$ over the whole $x$ regime by the appropriate Laplace transform of a L\'evy stable distribution $L_\mu$, that reads for the present isotropic two-dimensional problem \cite{Taqqu} :
\be
\Gene_{L}=\tilde{\mathcal{F}}_{L}(x)=\int_0^\infty {\rm d}z \, e^{-z - C(xz)^{\mu/2}}
\ee
where $\mu < 2$ and $C$ is a numerical constant. It is easy to check that for small $x$, $\tilde{\mathcal{F}}_{L}(x) \approx 1 - C \, \Gamma(1+\frac{\mu}{2})\, x^{\mu/2}$. Our main result is that this function is a very good fit to the empirical data corresponding to $\phi \approx \phi_J$ and $\tau \approx \tau^*$, for {\it all} values of $x$, see the bottom-left panel in Fig.~\ref{fig:gene}. The optimal value of $\mu$ is slightly smaller, $\mu \approx 1.6$, than the one obtained by a fit of the small $x$ region. The quality of the fit suggests that the distribution of displacements is indeed an isotropic L\'evy stable distribution $L_\mu$ of index $\mu$. 

We have checked that other distribution functions with power-law tails achieve a much poorer fit to the data. For example, a Student distribution of displacements 
(which cannot be obtained as the sum of independent jumps, contrarily to the L\'evy distribution) leads to:
\begin{align}
\Gene_{S} & = \tilde{\mathcal{F}}_{S}(x)\\
 & = 1 - (C'x)^{\mu/2} \int_0^\infty {\rm d}z \,\frac{e^{-z}}{(z + C'x)^{\mu/2}},\notag
\end{align}
where $\mu < 2$ and $C'$ is another numerical constant. Although $\Gene_{S}$ has the same small $x$ singularity and the same large $x$ behavior as $\Gene_{L}$, it turns out not to be possible to adjust the parameter $C'$ in such a way to adjust simultaneously the small and large $x$ behavior of the empirical $\Gene$. 

What is most significant, however, is that the value of the L\'evy index $\mu$ corresponding to the best fit turns out to be very close to the one expected in the case of a L\'evy flight where the diffusion exponent $\nu$ is given by $1/\mu$, since we find $\nu \approx 0.65$ (see Fig.~\ref{fig:rms-dv}) whereas $1/\mu \approx 0.625$.  

Fig.~\ref{fig:gene} bottom-right gives the value of the fitting parameter $\varepsilon = 1 - \mu/2$ in the plane $\phi,\tau$, obtained from the small $x$ behavior of $\Gene$, corresponding to large displacements. We see that the effective values of $\varepsilon$ (resp. $\mu$) become closer to $0$ (resp. $2$), corresponding to normal diffusion, as when $|\phi - \phi_c|$ and/or $|\ln \tau/\tau^*|$ increase. We believe that this corresponds to a so-called ``truncated'' L\'evy flight \cite{Koponen}, with a cut-off value in the distribution of elementary jump size that becomes smaller and smaller as one departs from the critical point, rather than a continuously varying exponent $\mu$. In other words, a plausible scenario that explains the behavior of $\varepsilon$ as a function of $\phi,\tau$ is that the tail of the elementary jump size distribution is given by:
\be 
P_1(r) \sim \frac{[r_0(\phi)]^\mu}{r^{1+\mu}},
\ee
where the exponent $\mu \approx 1.6$ is independent of $\phi$, whereas the typical scale of the jumps $r_0(\phi)$ and the cut-off $r_{\max}(\phi)$ both depend on $\phi$. It is reasonable that $r_{\max}(\phi)$ only diverges at $\phi=\phi_J$, while $r_0(\phi)$ has a regular, decreasing behavior as a function of $\phi$. In fact, the parameter $C$ appearing in the L\'evy distribution above is 
directly proportional to $r_0^\mu$. 

Assuming that the jumps are independent, the distribution of displacements $P_\tau(r)$ on a time scale $\tau$ is given by the $\tau$-th convolution of $P_1(r)$, which converges to a L\'evy distribution of order $\mu$ when $r_{\max} = \infty$ and predicts $r \propto \tau^\nu$ with $\nu=1/\mu$ (see e.g. \cite{BG}). For finite $r_{\max}$, $P_\tau$ will be very close to $L_\mu$ in the intermediate regime $1 \ll \tau \ll \tau_D$, where $\tau_D \sim (r_{\max}/r_0)^{\mu}$, before crossing over to a diffusive regime at very long times.  If $r_{\max}(\phi)$ behaves as $|\phi-\phi_J|^{-\zeta}$, the diffusion time $\tau_D$ should diverge as $|\phi-\phi_J|^{-\zeta\mu}$ when approaching $\phi_J$. 

\begin{figure}[t!]
\centering
\includegraphics[width=0.60\columnwidth]{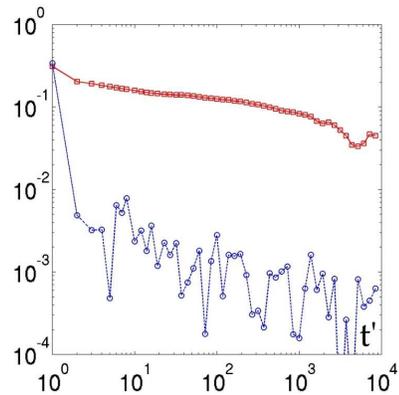}
\caption{
Correlation of the instantaneous ($\tau = 1$) velocity vectors $|\langle \br_i^{t}.\br_i^{t+t'} \rangle_{i,t}|$ (blue circles) versus the correlation of their amplitudes $\langle r_i^t.r_i^{t+t'}\rangle_{i,t}$ (red squares) as functions of the lag $t'$, for $\phi \approx \phi_J$.
}
\label{fig:corr}
\end{figure}

In order to check directly whether the jumps are indeed independent, we have measured the correlation of instantaneous ($\tau = 1$) velocities $\langle \br_i^{t}.\br_i^{t+t'} \rangle_{i,t}$. As
shown in Fig.~\ref{fig:corr}, this correlation function indeed decays extremely fast with the lag $t'$, excluding that superdiffusion could be due to long-range correlations in the displacements.
However, we discovered that the {\it amplitude} of the displacements, proportional to the parameter $r_0$, reveal long-range correlations, decaying approximately as $-\ln t'$. This suggests that some slow evolution takes place during an experimental run, that affects the value of $r_0$ over time. A natural mechanism is through the fluctuations of the local density $\phi$ (see below), that feedback on $r_0$. The approximately logarithmic decay of the correlations is interesting in itself and characteristic of multi-time scale glassy relaxation. These long-ranged correlations
slow down but fortunately do not jeopardize the convergence towards a L\'evy stable distribution.\footnote{Technically, however, these logarithmic correlations suggest that our superdiffusion process is a {\it multifractal L\'evy flight}, by analogy with the multifractal random walk introduced in \cite{MRW}. We will not dwell on this interesting new process, that would deserve a theoretical study of its own.} Therefore, the above interpretation in terms of a jump dominated superdiffusion, instead of persistent currents, is warranted.

\section{Dynamical correlations}

The above analysis focused on single grain statistics and established that the displacement of a grain has a L\'evy stable distribution in the superdiffusive regime. This means that the motion of each grain can be decomposed in a succession of power-law distributed jumps. But if ``large'' jumps can occur in such a jammed system, this necessarily implies that these jumps are correlated in space. In order to characterize these dynamical correlations, we have introduced in~\cite{Lech1} the following local correlation function:  
\be
q_i^t\left(a,\tau\right)\equiv \exp{-\frac{{r}_i^{t\, 2}\left(\tau\right)}{2a^2}},
\ee
where $a$ is a variable length scale over which we probe the motion. Essentially, $q_i^t\left(a,\tau\right) = 0$ if within the time lag $\tau$ the particle has moved more than $a$, and $q_i^t\left(a,\tau\right) = 1$ otherwise: $q_i^t\left(a,\tau\right)$ is a mobility indicator. The average over all $i$ and $t$ defines a function akin to the self-intermediate scattering function in liquids:
\be
\bar{Q}\left(a,\tau\right) \equiv \langle \frac1N \sum_i q_i^t\left(a,\tau\right)\rangle_{t}.
\ee
Note that it is closely related to the previously introduced generating function through $\bar{Q}\left(a,\tau\right)=\mathcal{F}\left(\lambda=\frac{1}{2a^2},\tau\right)$, which we have characterized in the previous section. 

\begin{figure}[t!]
\centering
\includegraphics[width=0.47\columnwidth]{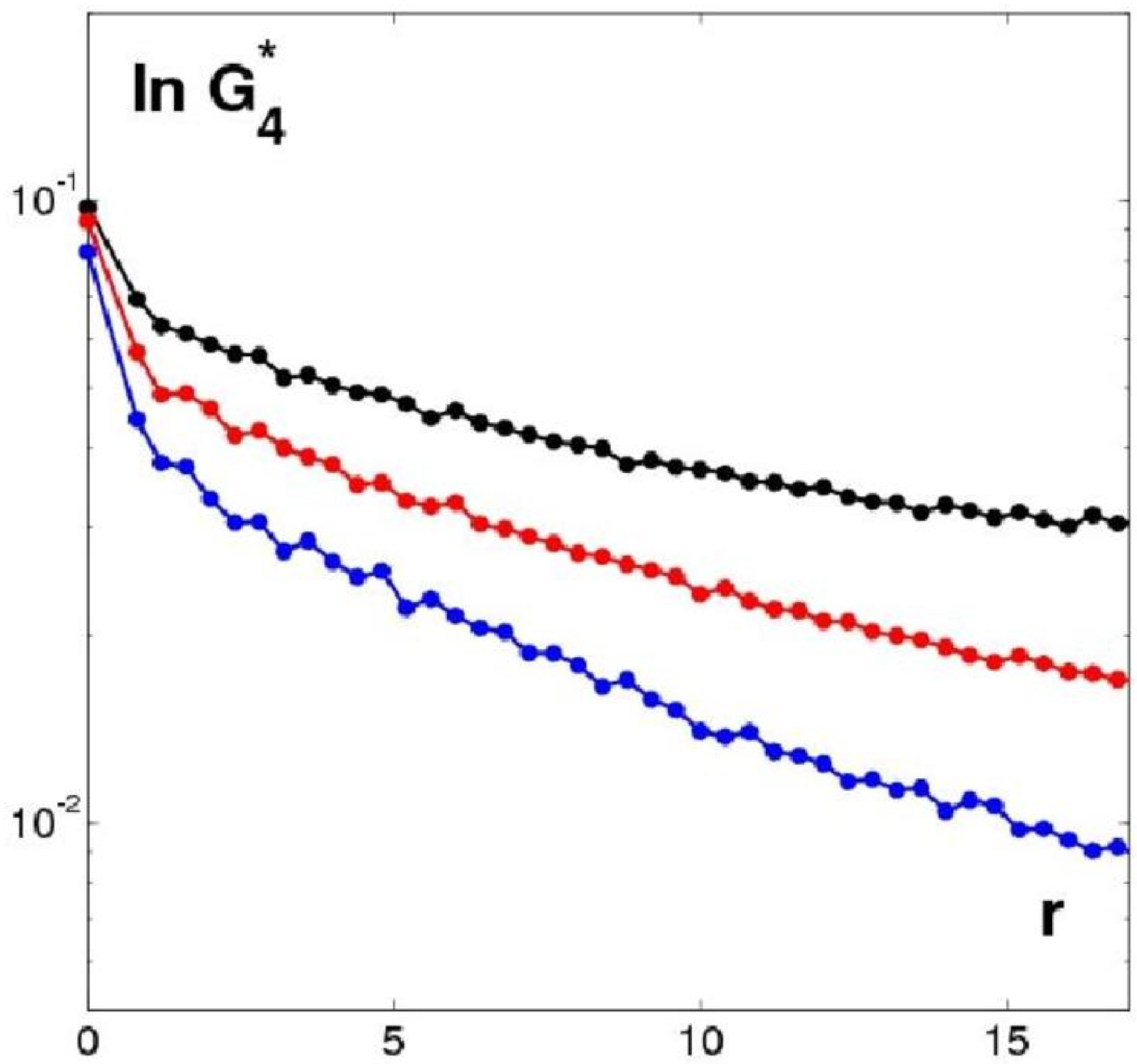}
\includegraphics[width=0.47\columnwidth]{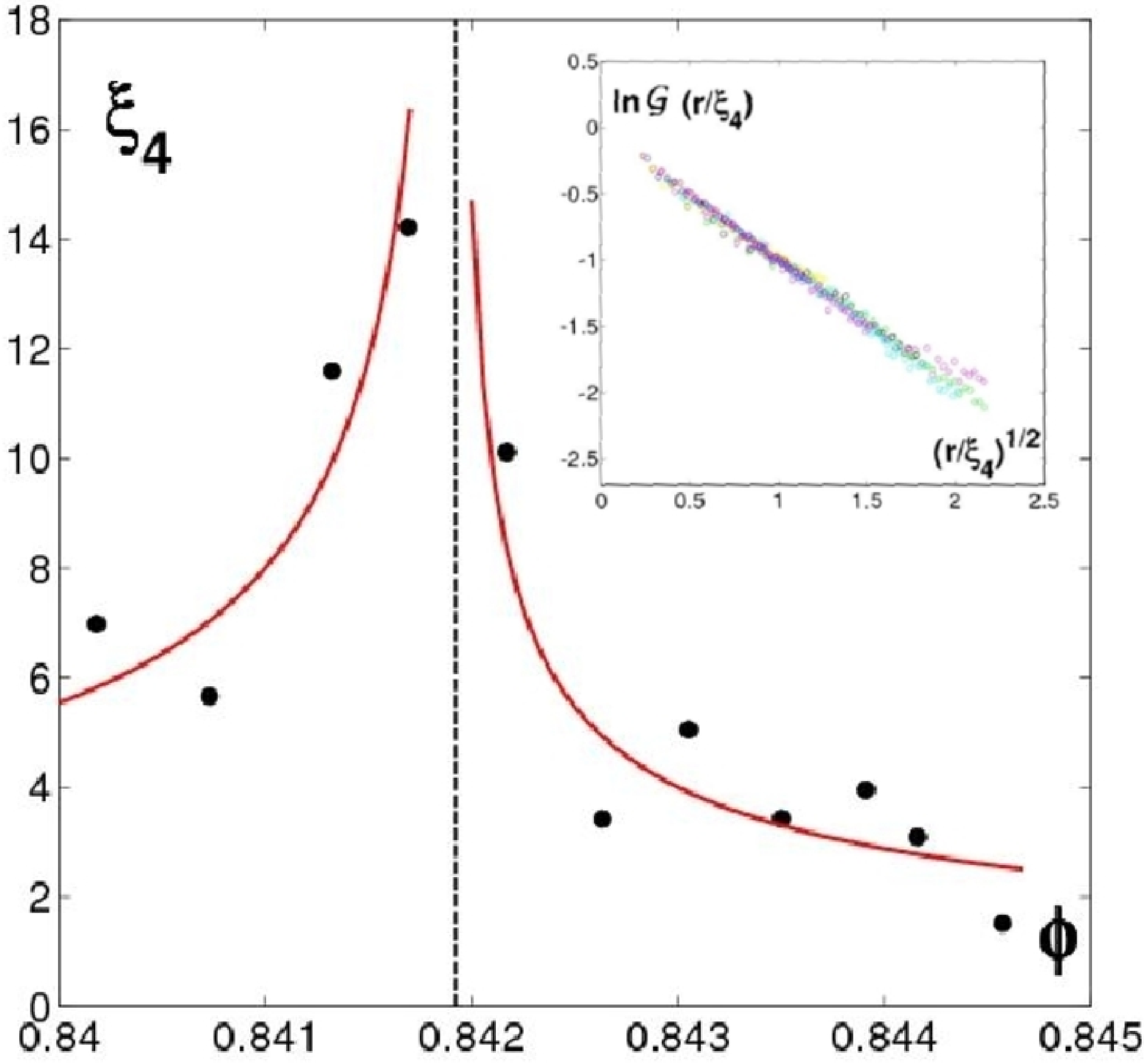}
\caption{
\leg{Left:} Four-point correlator $G_4^*(R)=G_4\left(R;\sigma_{\phi}\left(\tau^*\right),\tau^*\right)$ as a function of $R$ for $\phi = 0.8402$ (red), $0.8417$ (black), $0.8426$ (blue).
\leg{Right:} Dynamical correlation as a function of the packing fraction; (Inset: Rescaling of $\log\left[G_4^*(R)/G_4^*(0)\right]$ as a function of $\sqrt{R/\xi_4}$ for 8 densities around $\phi_J$).
From \cite{Lech1}.
}
\label{fig:G4}
\end{figure}

The dynamical (four-point) correlation function is defined as the spatial correlation of the $q$ field:
\be
G_4(R,a,\tau) = \left.\langle \delta q_i^t\left(a,\tau\right) \delta q_j^t\left(a,\tau\right)\rangle\right|_{t;|\vec R_i - \vec R_j|=R} 
\ee
where $\delta q_i^t\left(a,\tau\right)=q_i^t\left(a,\tau\right) - \bar{Q}\left(a,\tau\right)$ and is plotted for three packing fractions on fig.~(\ref{fig:G4})-left. Its sum over all $R$'s defines the so-called dynamical susceptibility $\Xifour$. Through simple manipulations, it is easy to show that $\Xifour$ is related to the variance of $Q$ as (see \eg \cite{JCP1} for a review):
\be
\Xifour = \frac1N \left\langle \left(\sum_i \delta q_i^t\left(a,\tau\right) \right)^2 \right\rangle_t.
\ee
We have shown in~\cite{Lech1} that $\Xifour$ has, for a given $\phi$, an absolute maximum $\chi_4^{*}=\chi_4(a^*,\tau^*)$, which as expected sits on the line corresponding to $\bar{Q}\left(a,\tau\right) \approx 1/2$, \ie such that half of the particles have moved by more than $a$. The amplitude 
of this maximum, which can be interpreted as a number of dynamically correlated grains, grows as $\phi$ approached $\phi_J$, indicating that the jump motion of the grains becomes more and more collective as one enters the superdiffusive L\'evy regime, as anticipated above. Note in particular that $\tau^*$ behaves in the same way as $\tau_{sD}$, the time at which the superdiffusion exponent $\nu$ reaches its maximum. We furthermore found~\cite{Lech1} that the four-point correlation $G_4^*(R) \equiv G_4(R,a^*,\tau^*)$ is a scaling function of $R/\xi_4$ (see inset of Fig.~\ref{fig:G4}-right), where $\xi_4(\phi)$ is the dynamical correlation length such that $\chi_4^{*} \propto \xi_4^2$.

All these results were reported in~\cite{Lech1} and are recalled here for completeness. The behavior of $\chi_4^*(\phi)$ was furthermore shown in~\cite{Lech2} to be well accounted for by the following upper bound, derived in~\cite{science,JCP1}: $\chi_{4} \ge ({\partial \bar{Q}}/{\partial \phi})^2 \langle \phi ^2 \rangle_c$, where $\langle \phi ^2 \rangle_c$ is the variance of the local density fluctuations. Here, we want to present further speculations, first on the relation between the size of the `micro-jumps' and dynamic correlations, and then on the higher moments of the distribution of $Q\left(a,\tau\right)$.

By conservation of the number of particles, the local change of density $\delta \phi$ is related to the divergence of the displacement field by: $\delta \phi/\phi = \vec \nabla \cdot \vec r$. If we invoke a kind a Reynolds dilatancy criterion whereby the local density must fall below some threshold for the system to move, the displacement field must be correlated over some length $\xi_4$ such that $\vec \nabla \cdot \vec r \sim r/\xi_4 \approx c$, where $c$ is a small number, possibly dependent on $\phi$, and of the order of $10^{-3} - 10^{-2}$ (\ie the relative difference between $\phi_J$ and $\phi_g$). This immediately leads to a relation between the typical size of the jumps after time $\tau$ and the required scale of the cooperative motion:
\be
\xi_4(\tau) \sim \frac{r(\tau)}{c} \propto \tau^\nu, \qquad \tau \leq \tau^*
\ee
This very simple argument predicts that $\xi_4$ should be $\sim 10^2 - 10^3$ times larger than typical displacements, which is indeed the case (see Fig.~\ref{fig:G4}). Furthermore, using $\tau^* \propto \tau_{D}$ one finds $r(\tau^*) \propto r_{\max}$, and therefore, using the truncated L\'evy flight model above, a power-law divergence of $\xi_4(\phi)$ as $|\phi-\phi_J|^{-\zeta}$. This divergence might change if $c$ strongly depends on the distance $|\phi-\phi_J|$.

Turning now to higher cumulants of the distribution of $Q\left(a^*,\tau^*\right)$, one can show that they are related to the space-integral of higher order correlation functions of the dynamical activity. For example, the skewness $\varsigma_6$ of $Q$ is $1/N^2$ times the space integral of the 6-point correlation function, defined as:
\be
G_6(R,R',a^*,\tau^*) = \left.\langle \delta q_i^{t*} \delta q_j^{t*} \delta q_k^{t*} \rangle \right|_{t;|\vec R_i - \vec R_j|=R; |\vec R_i - \vec R_k|=R'},
\ee
where $\delta q_i^{t*}$ is a shorthand notation for $\delta q_i^t\left(a^*,\tau^*\right)$.

Similar relations hold for higher moments. The simplest scenario is that all these higher order correlation functions are governed by the same dynamical correlation length $\xi_4$, extracted from $G_4(R,a,\tau)$. This is what happens in the vicinity of standard phase transitions, for example. If this is the case, and provided that $\chi_4^* \propto \xi_4^2$, one can show that the following scaling relations should hold:
\be\label{cumulants}
\varsigma_6 \sim \varsigma_{6,c} \sqrt{\frac{\chi_4^*}{N}}, \qquad \kappa_8 \sim \kappa_{c,8} \frac{\chi_4^*}{N}
\ee
where $\varsigma_6, \kappa_8$ are respectively the skewness and kurtosis of $1/N \sum_i q_i$ (related to the 6- and 8-point connected correlation functions), and $\varsigma_{c,6}, \kappa_{c,8}$ the corresponding values at the critical point, \ie for systems of size $N$ smaller than the correlation volume $\chi_4^* \propto \xi_4^2$. If these scaling relations are valid, the determination of $\varsigma_{c,6}, \kappa_{c,8}$ using the above equations should give the same values for any $\phi$ close to $\phi_J$. Unfortunately, our statistics is not sufficient to make definitive statements, although the data is indeed compatible with such scalings. The notable feature is that both $\varsigma_{c,6} \simeq -1$ and $\kappa_{c,8} \simeq -5$ are found to be {\it negative}, although the error
bar on both quantities is large. Interestingly, if we assume that the displacements are perfectly correlated within a correlation blob of size $\xi_4$, the L\'evy flight model with $\mu=1.6$ makes the following predictions:
\be
\varsigma_{c,6} \approx -0.12; \qquad \kappa_{c,8} \approx -1.37,
\ee
in qualitative agreement with our data (note however that there is an unknown proportionality factor in eq.~(\ref{cumulants}). A Gaussian diffusion model, on the other hand, predicts $\varsigma_{c,6}=0$
and $\kappa_{c,8} = -6/5$, corresponding to a uniform distribution of $Q$ in $[0,1]$. A kurtosis smaller than $-6/5$ means that the distribution of $Q$ tends to be spiked around $0$ and $1$. 
Finally, as $\phi$ increases beyond $\phi_J$, the negative skewness of $Q$ markedly increases. This is a sign that the dynamics becomes more and more intermittent, ``activated'', with a few rare events decorrelating the system completely, while most events decorrelate only weakly. Since we fix $\bar{Q}\left(a^*,\tau^*\right)=1/2$, this leads to a diverging negative skewness in the limit where the probability of rare decorrelating events tends to zero. The idea of characterizing the skewness of $Q$ might actually be an interesting tool to characterize the strength of activated events in other 
glassy systems.

\section{Conclusion}

The central finding of this work is that the superdiffusive motion of our frictional grains in the vicinity of the jamming transition appears to be a genuine L\'evy flight, but with `jumps' taking 
place on a Lilliputian scale. This interpretation contrasts with the original suggestion made in \cite{Lech1}, where anomalous diffusion was ascribed to persistent, spatially correlated 
currents. The vibration of the plate induces a broad distribution of jumps that are random in time, but correlated in space, and that can be interpreted as {\it micro-crack events} on all scales. 
As the volume fraction departs from the critical jamming density $\phi_J$, this distribution of jumps is truncated at smaller and smaller jump sizes, inducing a crossover towards standard diffusive motion
at long times. 
This picture severely undermines the usefulness of harmonic modes as a way to rationalize the dynamics of our system (although this conclusion might of course not carry over to frictionless grains, or 
to thermal systems). The detailed study of these modes, and the difficulty to analyze them in the present system, is discussed in \cite{companion}. 

We have also presented several other speculations about the relation between the dynamical correlation length $\xi_4$ and the size of the jumps, and the structure of higher order dynamical cumulants. 
The idea of using the 6-point skewness as a quantitative measure of the importance of activated events in the dynamics of glassy systems seems to us worth pursuing further. 

\vskip 0.5cm
{\it Note added and acknowledgment:} We want to thank L. Berthier for interesting discussions about a preprint of his and collaborators \cite{Berthier} (that appeared in the very last stages of the present work), and where effects similar to those discussed here are observed numerically in a model system. We thank him in particular for insisting on the importance of measuring velocity correlations in our
system.

\end{document}